\documentclass[12pt]{article}
\usepackage{amsfonts,amssymb,amsmath}
\usepackage[dvips]{epsfig}
\begin{document}
\title{\bf\large{Minimum error discrimination between similarity transformed quantum states}}\vspace{20mm}
\author{M. A. Jafarizadeh $^{a,b,c}$
 \thanks{E-mail:jafarizadeh@tabrizu.ac.ir},
 R. Sufiani$^{a,b}$
 \thanks{E-mail:sofiani@tabrizu.ac.ir}, Y. Mazhari $^{a}$ \thanks{E-mail:Mazhari@tabrizu.ac.ir}
\\
$^{a}${\small Department of Theoretical Physics and
Astrophysics, University of Tabriz, Tabriz 51664, Iran.}  \\
$^b${\small Institute for Studies in Theoretical Physics and
Mathematics, Tehran 19395-1795, Iran.}\\$^{c}${\small Research
Institute for Fundamental Sciences, Tabriz 51664, Iran.}}
\pagebreak
 \newtheorem{thm}{Theorem}
 \newtheorem{cor}[thm]{Corollary}
 \newtheorem{lem}[thm]{Lemma}
 \newtheorem{prop}[thm]{Proposition}
 \newtheorem{defn}[thm]{Definition}
 \newtheorem{rem}[thm]{Remark}
\vspace{20mm}

\maketitle

\begin{abstract}
Using the known necessary and sufficient conditions for minimum
error discrimination (MED), first it is shown that a Helstrom
family of ensembles is equivalent to these conditions and then by
a convex combination of the initial states (the states which we
try to discriminate them) and the corresponding conjugate states,
a more suitable and convenient form for the MED conditions is
extracted, so that optimal set of measurements and corresponding
optimal success probability of discrimination can be determined.
Then, using the introduced identity, MED between $N$ similarity
transformed equiprobable quantum states is investigated. As a
special case, MED between the so called group covariant or
symmetric states
is considered.\\

{\bf Keywords: Minimum error discrimination (MED), Helstrom family
of ensembles, similarity transformed quantum states, POM, group
covariant states}

{\bf PACs Index: 01.55.+b, 02.10.Yn }
\end{abstract}
\section{Introduction}
The theory of quantum information and communication concerns the
transmission of information using quantum states and channels. The
transmission party encodes a message onto a set of quantum states
$\rho_i$ with prior probability $p_i$ for each state $\rho_i$. The
task of the receiving party is to decode the received message,
i.e., finding the best measurement strategy based upon the
knowledge of the signal states and their prior probabilities. One
possibility is to choose the strategy that minimizes the
probability of error. In order that a set of probability operator
measure (POM) minimizes the error probability of a detection, it
must satisfy a known set of necessary and sufficient conditions
\cite{Helstrom}-\cite{Yuen}. Using the no-signaling principle, an
upper bound to the success probability in the minimum-error state
discrimination has been given \cite{Hwang}. The problem of minimum
error probability discrimination of symmetric quantum states has
been studied in \cite{Assalini,Barnett2,Chou}. For $N$ symmetric
pure states occurring with equal prior probabilities, the optimal
minimum error measurement is the square root measurement (SRM)
\cite{Barnett3}. In this paper, first we show that the known
necessary and sufficient conditions for MED \cite{Barnett1} are
equivalent to a Helstrom family of ensembles \cite{Helstrom},
i.e., the optimal conditions for MED are fulfilled by a Helstrom
family of ensembles. Then, we extract a suitable identity from the
MED conditions
 which is more convenient for the study of mixed quantum states
 discrimination with optimal success probability. By using the
 introduced identity, we study MED between $N$ equiprobable similarity transformed qudit
states (quantum states in $d$ dimensional Hilbert space)
$\rho_{1},\ldots,\rho_{N}$, defined by $\rho_{i} =U_i\rho_1
U_i^{-1}$, so that the unitary operators $U_1\equiv I_d,
U_2,\ldots, U_N$ generate a (finite or continuum) subgroup of the
unitary group $U(d)$. It is shown that, in the case in which
$U_i$'s generate a non-abelian subgroup, the state space is
irreducible (none of the states have invariant components under
the action of $U_ i$) and the optimal discrimination can be
achieved. In the case that $U_i$'s generate an abelian subgroup,
for instance rotations about a fixed axis, the state space is
reducible and although there is no closed-form formula   in
general case but the procedure can be applied in each case in
accordance to that case.
\section{Minimum error discrimination between
quantum states and Helstrom family of ensembles} In general, the
measurement strategy is described in terms of a set of
non-negative-definite operators called the probability operator
measure (POM). The measurement outcome labeled by $i$ is
associated with the element $\Pi_i$ of POM that has all the
eigenvalues either positive or zero. The POM elements must add up
to the identity operator, i.e., $\sum_i\Pi_i={\mathbf{1}}$. The
probability that the receiver will observe the outcome $i$ given
that the transmitted signal is $\rho_j$ is given by $p(i|j)=Tr
(\Pi_i\rho_j)$. Assume that given states
$\rho_{1},\rho_{2},\ldots,\rho_{N}$ have prior probabilities
$p_{1},p_{2},\ldots,p_{N}$, respectively $(p_{i}\geq
0,\sum_{i}p_{i}=1)$. It follows that the probability for correctly
identifying states $\rho_i$ is given by
\begin{equation}\label{e2}
p=1-p_{error}=\sum^{N}_{i=1}p_{i}Tr(\rho_{i}\Pi_{i}),
\end{equation}
where $p_{error}$ is the error probability.

The necessary and sufficient condition that lead to the
minimum-error probability is known to be \cite{Barnett1}
\begin{equation}\label{e8}
\sum^{N}_{i=1}p_{i}\Pi_{i}\rho_{i}-p_{j}\rho_{j}\geq0\quad,
\forall \;\ j=1,...,N.
\end{equation}
While these conditions do give us a starting point for finding
minimum error POM's, they do not themselves provide a great
insight into either the form of minimum error measurement
strategies, or into how error probability depends on the set of
possible states. For this we must examine the solutions to these
conditions, and there are not many solutions. The essential
difficulty in solving the conditions directly is that all of the
variables $\Pi_k$ appear in each condition, and they are not
independent variables.

In the following we use the fact that, the inequality (\ref{e8})
indicates
\begin{equation}\label{e10}
\sum^{N}_{i=1}p_{i}\rho_{i}\Pi_{i}-p_{j}\rho_{j}=\alpha_{j}\tau_{j},\;\;\;\
\forall\;\ j=1,\ldots, N.
\end{equation}
where $\alpha_{j}\geq0$ and $\tau_j$'s are positive operators. By
taking the trace of both sides of Eqs. (\ref{e10}) and using
(\ref{e2}), we obtain $\alpha_{j}=p_{opt}-p_{j}$ (In
\cite{hels.maz} it has indicated that $p_{opt}\geq p_{j}$). If we
postmultiply Eq. (\ref{e10}) by $\Pi_{j}$ and summing up over $j$,
we get $\sum^{N}_{j=1}(p_{opt}-p_{j})\tau_{j}\Pi_{j}=0$. Because
both $\tau_{j}$ and $\Pi_{j}$ are positive operators it follows
that $(p_{opt}-p_{j})\tau_{j}\Pi_{j}=0$ for every $j=1,\ldots, N$.
This indicates that $\tau_{j}\Pi_{j}=0$ for all $j=1,\ldots, N$.
Moreover, in order that the optimal measurement operators $\Pi_j$
can be constructed, the states $\tau_j$ must be possess at least
one zero eigenvalue, i.e., $\tau_j$s are not full rank. By
denoting the term $\sum^{N}_{i=1}p_{i}\rho_{i}\Pi_{i}$ by
$\mathcal{M}$, the necessary and sufficient conditions for
minimum-error probability, i.e., the conditions (\ref{e10}) take
the following form
\begin{equation}\label{e100}
\mathcal{M}=p_{j}\rho_{j}+(p_{opt}-p_j)\tau_{j},\;\;\;\ \forall\;\
j.
\end{equation}
For the case that all given states $\rho_{i}$, $i=1,\ldots, N$
have equal prior probability $p_i=\frac{1}{N}$, we have
\begin{equation}\label{e100'}
\mathcal{M}=\frac{1}{N}\rho_{j}+(p_{opt}-\frac{1}{N})\tau_{j},\;\;\;\
\forall\;\ j.
\end{equation}
Although, we will use the identity (\ref{e100}) - equivalent to
necessary and sufficient conditions (\ref{e8}) - in order to
discriminate quantum states with minimum error, in the following
we show that a Helstrom family of ensembles is necessary and
sufficient for realizing a minimum error measurement, i.e., the
necessary and sufficient conditions (\ref{e8}) are fulfilled by a
Helstrom family of ensembles.
\subsection{Equivalence between  optimality conditions and Helstrom
family of ensembles} First let us to recall the definition of a
Helstrom family of ensembles. A set of $N$-numbers
$\{\tilde{p}_{i},\rho_{i};1-\tilde{p}_{i},\tau_{i}\}^{N}_{i=1}$ is
called a weak Helstrom family (of ensembles) if there exist
N-numbers of binary probability discriminations
$\{\tilde{p}_{i},1-\tilde{p}_{i}\}^{N}_{i=1}(0<\tilde{p_{i}}\leq1)$
and states $\{\tau_{i}\in\emph{s}\}^{N}_{i=1}$ satisfying
\begin{equation}\label{e3}
p=\frac{p_{i}}{\tilde{p}_{i}}\leq 1
\end{equation}
and
\begin{equation}\label{e4}
p_{i}\rho_{i}+(p-p_{i})\tau_{i}=p_{j}\rho_{j}+(p-p_{j})\tau_{j},
\end{equation}
for any $i,j=1,\ldots,N$ \cite{Kimura}. We assume that a prior
probability distribution satisfies $p_{i}\neq 0,1$ in order to
remove trivial cases. $p$ and $\tau_{i}$ are called  Helstrom
ratio and conjugate state to $\rho_{i}$, respectively. Multiplying
(\ref{e4}) with $\Pi_{i}$, taking the sum over $i$ and then taking
the trace of both sides leads to
\begin{equation}
Tr\mathcal{M}+\sum^{N}_{i=1}(p-p_{i})Tr(\tau_{i}\Pi_{i})=p.
\end{equation}
Thus we have
\begin{equation}\label{e5}
p_{opt}\leq p.
\end{equation}
The observables $\{\Pi_{i}\}^{N}_{i=1}$ satisfy $p_{opt}=p$ if
$(p-p_{i})Tr(\tau_{i}\Pi_{i})=0$, $i=1,\ldots,N$. In this case,
the observables $\{\Pi_{i}\}^{N}_{i=1}$ give an OM to
discrimination of states $\{\rho_{i}\}^{N}_{i=1}$, and we call the
family
$\{\tilde{p}_{i},\rho_{i};1-\tilde{p}_{i},\tau_{i}\}^{N}_{i=1}$
Helstrom family of ensembles \cite{Kimura}.

Now, we show the equivalence between optimality conditions
(\ref{e8}) and Helstrom family of ensembles. To see that Helstrom
family of ensembles is sufficient to minimize the error let us
postmultiply (\ref{e4}) by $\Pi_{i}$. If we consider
$(p-p_{i})\tau_{i}\Pi_{i}=0$ then
\begin{equation}\label{e9}
p_{i}\rho_{i}\Pi_{i}=p_{j}\rho_{j}\Pi_{i}+(p-p_{j})\tau_{j}\Pi_{i}
\end{equation}
Summing up over $i$ and using the completeness condition
$\sum_{i=1}\Pi_i=I$ for probability operators, we obtain
\begin{equation}\label{e10x}
\mathcal{M}=p_{j}\rho_{j}+(p-p_{j})\tau_{j},\;\;\ \forall \;\ j
\end{equation}
which is the same condition (\ref{e8}) or equivalently
(\ref{e100}).

To show that Helstrom family of ensembles is necessary condition,
we take the trace of Eqs. (\ref{e10}) and obtain
$\alpha_{j}=p-p_{j}$. Then, by subtracting Eqs. (\ref{e10}) for
different values of $j$, the Eq. (\ref{e4}) is followed. If we
postmultiply Eq. (\ref{e10x}) by $\Pi_{j}$ and summing up over
$j$, we get just $\sum^{N}_{j=1}(p-p_{j})\tau_{j}\Pi_{j}=0$.
Because both $\tau_{j}$ and $\Pi_{j}$ are both positive operators
it must then be the case that $(p-p_{j})\tau_{j}\Pi_{j}=0$.
\section{Minimum error discrimination (MED) between similarity transformed quantum states}
Let $\{U_1=I_d,U_2,\ldots, U_{N}\}$ be a generating set for a
(finite or continuum) subgroup of the unitary group $U(d)$. Now,
we consider the states $\rho_i$ and the corresponding conjugate
states $\tau_i$ as
\begin{equation}\label{rs}
\rho_{i} =U_i\rho_1 U_i^{-1},\;\;\ \tau_{i}=U_i\tau_1
U_i^{-1},\;\;\ \forall \;\ i=1,2,\ldots, N.
\end{equation}
We will refer to such states as similarity transformed states.
Then, we have
$$\mathcal{M}=U_i[\frac{1}{N}\rho_1 +(p-\frac{1}{N})\tau_{1}]U_i^{-1}=U_i\mathcal{M} U_i^{-1},\;\ i=1,\ldots,  N.$$
This indicates that $\mathcal{M}$ commutes with the generators
$U_i$, and so commutes with all elements of the subgroup $G$
generated by $U_i$, i.e., we have
\begin{equation}\label{m}U(g)\mathcal{M}=\mathcal{M}U(g), \;\;\ \forall \;\ g\in
G,
\end{equation}
where, $U$ is a $d\times d$ unitary representation of $G$. A
consequence of the above equation is that $\mathcal{M}$ is a
diagonal matrix. To see this, we first consider that the
representation $U$ be irreducible ($U$ is called as irreducible if
it has no non-trivial - other than $\{0\}$ and the hole Hilbert
space- invariant subspaces). Then, the well known Schur's first
lemma in the group representation theory \cite{joshi} implies that
$\mathcal{M}$ is a constant multiple of the identity matrix, i.e.,
$\mathcal{M}=\mu I_d$ for some complex number $\mu\in
\mathcal{C}$. If $U$ be reducible, one can decompose it as
 nonequivalent irreducible components $V_i$. For instance, assume
 that $U$ be a direct sum of two nonequivalent irreducible
 representations $V_1$ and $V_2$. Then, the commutativity relation
 (\ref{m}) can be written as
 $$U(g)\mathcal{M}=\left(\begin{array}{cc}
   V_1(g) &  \\
      & V_2(g) \\
   \end{array}\right)\left(\begin{array}{cc}
   A &  B\\
    C  & D \\
   \end{array}\right)=\left(\begin{array}{cc}
   A & B \\
    C  & D \\
   \end{array}\right)\left(\begin{array}{cc}
   V_1(g) &  \\
      & V_2(g) \\
   \end{array}\right)=\mathcal{M}U(g),\;\ \forall \;\ g\in G,
$$
which indicates that \begin{equation}\label{u}V_1(g)A=AV_1(g),\;\
V_2(g)D=DV_2(g), \;\ V_1(g)B=BV_2(g),\;\ V_2(g)C=CV_1(g),\;\
\forall \;\ g\in G.
\end{equation}
Now, from the Schur's first lemma, we conclude that $A=\alpha I_1$
and $D=\beta I_2$ ($I_1$ and $I_2$ are identity matrices of
dimension $dim V_1$ and $dim V_2$, respectively). From the
relations (\ref{u}), and the fact that $V_1$ and $V_2$ are
nonequivalent irreducible representations, the Schur's second
lemma implies that $B=C=\mathbf{0}$ where, $\mathbf{0}$ denotes
zero matrix. The proof for diagonality of $\mathcal{M}$ in general
case in which $U$ is decomposed to more than two irreducible
representations, is similar.

In order to $\Pi_i$'s form a set of POM, we need to have the
completeness condition $\sum_{i=1}^N\Pi_i=I$. Assume that
$\Pi_i=\lambda_i\Pi'_i$ for some positive complex number
$\lambda_i$ and positive operator $\Pi'_i$ so that, the condition
$\Pi_i\tau_i=0$ implies $\Pi'_i\tau_i=0$ for all $i=1,\ldots, N$.
The operators $\Pi'_i$ can be obtained from a positive operator
$\Pi'_1$ via the same similarity transform which defines the
states $\rho_i$ and the corresponding conjugate states $\tau_i$,
i.e., we consider
\begin{equation}\label{p}
\Pi'_i=U_i\Pi'_1U_i^{-1}, \;\;\ i=1,2,\ldots, N.
\end{equation}
Therefore, it is sufficient to choose $\Pi'_1$ perpendicular to
$\tau_1$ in order to have $\Pi'_i\tau_i=0$ for all $i$. Then, we
must have
\begin{equation}\label{p'}
\sum_{i=1}^N\Pi_i=\sum_{i=1}^N\lambda_i\Pi'_i=I,
\end{equation}
with $\sum_{i=1}^N\lambda_i=1$; That is, the convex hull of the
operators $\Pi'_i$, $i=1,\ldots, N$ must conclude the identity
operator $I$. Also, the completeness condition (\ref{p'}) leads to
the following fact
\begin{equation}\label{p''}
\sum_{i=1}^N\lambda_iTr(\Pi'_i)=\sum_{i=1}^N\lambda_iTr(\Pi'_1)=d
\;\ \rightarrow \;\ Tr(\Pi'_1)=d
\end{equation}
where, we have used the fact that
$Tr(\Pi'_i)=Tr(U_i\Pi'_1U_i^{-1})=Tr(\Pi'_1)$ for all $i$.

Now, by choosing suitable operator $\Pi'_1$ perpendicular to
$\tau_1$, and using (\ref{e2}), one can obtain the optimal success
probability of discrimination between quantum states $\rho_i$,
$i=1,2,\ldots, N$ with equal prior probability $p_i=\frac{1}{N}$,
as follows
\begin{equation}\label{prob}
p_{_{opt}}=\frac{1}{N}\sum_{i=1}^N\lambda_iTr(\Pi'_i\rho_i)=\frac{1}{N}\sum_{i=1}^N\lambda_iTr(\Pi'_1\rho_1)=\frac{1}{N}Tr(\Pi'_1\rho_1).
\end{equation}
The above result shows that the positive coefficients $\lambda_i$
in the convex combination (\ref{p'}) can be chosen arbitrarily in
a way that $\sum_i\lambda_i=1$, i.e., the optimal POM set
$\{\Pi_i=\lambda_i\Pi'_i, \;\ i=1,2,\ldots, N\}$ satisfying the
optimality condition (\ref{e10}) - provided that $\Pi'_i\tau_i=0$
- is not unique.\\\\
\textbf{A. The irreducible case}\\
In the case that the subgroup generated by $U_i$, $i=1,2,\ldots,
N$ is a non-abelian subgroup of $U(d)$, the only operator which
can be invariant under the action of representation $U$ is
multiple of identity operator. Therefore, from (\ref{m}) we have
$\mathcal{M}=\mu I$ and so the identity (\ref{e100'}) is written
as the following resolution of identity
\begin{equation}\label{e10'}
\mathcal{M}=\mu
I=\frac{1}{N}\rho_{i}+(p_{_{opt}}-\frac{1}{N})\tau_{i},\;\ \forall
\;\ i=1,\ldots, N.
\end{equation}
Taking the trace of both sides of (\ref{e10'}), we find
\begin{equation}\label{e10''}
d\mu=\frac{1}{N}+(p_{_{opt}}-\frac{1}{N})=p_{_{opt}} \;\;\
\rightarrow \;\ \mu=\frac{p_{_{opt}}}{d}.
\end{equation}
Now, in order to consider optimal discrimination between the
states $\rho_i$ in (\ref{rs}), we assume that
$\rho_1=\sum_{i=1}^da_i| \psi^{(1)}_i\rangle\langle \psi^{(1)}_i|$
be mixed state. Then, the resolution of identity (\ref{e10'})
implies that $\tau_1$ is also diagonal in the bases $|
\psi^{(1)}_i\rangle$. In the case that $\rho_1$ is full rank
($a_i\neq 0$ for all $i=1,2,\ldots,d$), the state $\tau_1$ can be
written as $\tau_1=\sum_ib_i|\psi^{(1)}_i\rangle\langle
\psi^{(1)}_i|$ so that at least one of the coefficients $b_i$ is
zero. Then, by using (\ref{e10'}) and (\ref{e10''}), we have
$$\frac{p_{_{opt}}}{d} (\sum_{i=1}^d|
\psi^{(1)}_i\rangle\langle
\psi^{(1)}_i|)=\frac{1}{N}[\sum_{i=1}^da_i|
\psi^{(1)}_i\rangle\langle
\psi^{(1)}_i|+(p_{_{opt}}N-1)\sum_{i=1}^db_i|
\psi^{(1)}_i\rangle\langle \psi^{(1)}_i|],$$ so that
$$\frac{p_{_{opt}}}{d}=\frac{1}{N}[a_i+(p_{_{opt}}N-1)b_i].$$
Since, at least one of the coefficients $b_i$, say $b_l$, is zero,
the above relation leads to the following result
\begin{equation}\label{res}
p_{_{opt}}=\frac{d}{N}a_{max}
\end{equation}
where, $a_{max}$ is the largest eigenvalue of $\rho_1$. From the
fact that $a_{max}\geq \frac{1}{d}$, it is seen that $p\geq
\frac{1}{N}$. It should be noticed that, in the case that all
$a_i$'s are distinct, only one of the coefficients $b_i$, say
$b_l$, must be zero so that $\Pi'_1$ will be given as
$\Pi'_1=d|\psi^{(1)}_l\rangle \langle \psi^{(1)}_l|$ and
consequently the set of POMs $\Pi_i$ are pure, i.e.,
$\Pi_1=\frac{d}{N}| \psi^{(1)}_l\rangle\langle \psi^{(1)}_l|$.

Now, let there be $m$ independent eigenvectors of $\rho_1$ having
the same maximum eigenvalue $a_{max}$. Then, $m$ eigenvalues of
$\tau_1$ must be zero. Denoting these eigenvalues by
$b_{i_1},\ldots, b_{i_m}$, the operator $\Pi'_1$ will be written
as $$\Pi'_1=\alpha_1|\psi^{(1)}_{i_1}\rangle \langle
\psi^{(1)}_{i_1}|+\ldots+\alpha_m|\psi^{(1)}_{i_m}\rangle \langle
\psi^{(1)}_{i_m}|,$$ where $\alpha_i$'s are arbitrary complex
numbers with $\sum_{i=1}^m\alpha_{i}=d$. Then, using (\ref{prob})
we obtain the same result (\ref{res}) for the optimal success probability.\\

For instance, in the case $d=2$, we consider
$\rho_i=U_i\rho_1U^{-1}_i$ with
$$\rho_1=\frac{1}{2}(I+a\hat{n}.\vec{\sigma})=\frac{1+a}{2}|+n\rangle\langle +n|+\frac{1-a}{2}|-n\rangle\langle -n|$$ with $|a|<1$, $|+n\rangle=\frac{1}{\sqrt{2}}\left(\begin{array}{c}
          1 \\
           e^{i\varphi}  \\
         \end{array}\right)$ and $|-n\rangle=\frac{1}{\sqrt{2}}\left(\begin{array}{c}
          1 \\
           -e^{i\varphi}\\
         \end{array}\right)$. Then, we have
$\tau_i=U_i\tau_1U^{-1}_i$ with
$\tau_1=\frac{1}{2}(I-\hat{n}.\vec{\sigma})=|-n\rangle\langle -n|$
and, the optimal success probability reads as
\begin{equation}\label{e11''}
p_{opt}=\frac{1+a}{N}.
\end{equation}
The above result has been obtained in Ref.\cite{hels.maz} via the
Helstrom family of ensemble.

Now, assume that $\rho_1$ has rank $r< d$, i.e., $a_i=0$ for
$i=r+1,\ldots, d$. Then, the completeness relation (\ref{e10'})
can be written as
$$\frac{p_{_{opt}}}{d} (\sum_{i=1}^r|
\psi^{(1)}_i\rangle\langle \psi^{(1)}_i|+\sum_{i=r+1}^d|
\phi^{(1)}_i\rangle\langle
\phi^{(1)}_i|)=$$$$\frac{1}{N}[\sum_{i=1}^ra_i|
\psi^{(1)}_i\rangle\langle
\psi^{(1)}_i|+(p_{_{opt}}N-1)(\sum_{i=1}^rb_i|
\psi^{(1)}_i\rangle\langle \psi^{(1)}_i|+\sum_{i=r+1}^d b_i|
\phi^{(1)}_i\rangle\langle \phi^{(1)}_i|)],$$ so that
$$\frac{p_{_{opt}}}{d}=\frac{1}{N}[a_i+(p_{_{opt}}N-1)b_i], \;\;\ i=1,\ldots, r$$
and $$\frac{p_{_{opt}}}{d}=\frac{1}{N}(p_{_{opt}}N-1)b_i, \;\;\
i=r+1,\ldots, d.$$ Therefore, we obtain the same result
(\ref{res}).

In the special case in which the initial state $\rho_1$ is pure,
i.e., we have $a_{i}=0$ for all $a_i$ except for one, say $a_l$,
which is equal to one, the result (\ref{res}) leads to the
following optimal success probability
\begin{equation}
p_{_{opt}}=\frac{d}{N}.
\end{equation}\\
\textbf{B. The reducible case}\\
In the case that the subgroup generated by $U_i$, $i=1,\ldots, N$
is an abelian subgroup of $U(d)$, invariance of $\mathcal{M}$
under the action of $U$ (Eq.(\ref{m})) implies that $\mathcal{M}$
is diagonal but not in general proportional to the identity
matrix. Therefore, in order to determine the set of optimal POM in
this case, one can consider that the state $\rho_1$ be diagonal,
so that the identity (\ref{e10''}) will be lead to the fact that
$\tau_1$ is also diagonal. Then, similar to the irreducible case,
one can obtain the suitable $\tau_1$ which fulfill the minimum
error condition (\ref{e10''}). Bellow, we consider an example in
details, in order to clarify the method.

Let us consider the abelian subgroup $SO(2)$ of the rotation group
$SO(3)$ which is generated by a rotation operator as
$\exp(-i\theta \hat{n}.\vec{J})$, that rotates a spin-$j$ state by
$\theta$ with respect to the $\vec{J}$-axis. Based on the rotation
picture, the ensembles of $\rho_{k}$ can be constructed as
follows. The states $\rho_{k}$ for $k = 1, \ldots ,N $, that we
wish to discriminate among are:
$$\rho_1=\frac{1}{d}(I+2a\hat{n}.\vec{J}),$$
\begin{equation}\label{x0}
\rho_{k}=U_k\rho_1 U_k^{-1},
\end{equation}
where $U_k=\exp{(\frac{2\pi i(k-1) J_z}{N})}$ is a rotation of
magnitude $\frac{2\pi (k-1)}{N}$ about the $z$-axis.
 The states $\tau_{k}$ are constructed similarly via
 $\tau_1=\frac{1}{d}(I+2b\hat{n}.\vec{J})$.

We note that in order to $\rho_1$ be a density matrix, its
eigenvalues $\lambda_m$ given by $\lambda_m=\frac{1+2am}{d}$ for
$-j\leq m\leq j$ must be non-negative, i.e., we have $a\leq
\frac{1}{2j}$. Also, it should be noticed that, $\tau_1$ is not
full rank and so, its minimum eigenvalue is zero which indicates
that $b=\frac{1}{2j}$.

Now, we want to obtain the upper bound $p$ for the optimal success
probability. To this end, we note that in this case, although
$\mathcal{M}$ is invariant under the action of operators $U_k$,
but it is not proportional to the identity matrix. In this case,
the invariance of $\mathcal{M}$ under rotations about the $z$-axis
implies that, it has the following form
\begin{equation}\label{x1}
\mathcal{M}=\alpha I_d+\beta J_z,
\end{equation}
where the constants $\alpha$ and $\beta$ must be calculated. Then,
for a given mixed state $\rho_1$ as in (\ref{x0}) and the
corresponding $\tau_e$ as
$$\tau_1=\frac{1}{d}(I+2b\hat{n'}.\vec{J}),$$
with $b=\frac{1}{2j}=\frac{1}{d-1}$, we have
$$
\mathcal{M}=\frac{1}{N}\{\rho_1+(Np-1)\tau_1\}=$$
\begin{equation}\label{x2}
\frac{1}{Nd}\{Np
I_d+2(an_x+(Np-1)bn'_x)J_x+2(an_y+(Np-1)bn'_y)J_y+2(an_z+(Np-1)bn'_z)J_z\}
\end{equation}
Comparing (\ref{x2}) by (\ref{x1}) results
\begin{equation}\label{x3}
\alpha=\frac{p}{d},\;\;\;\
\beta=2\frac{an_z+(Np-1)bn'_z}{Nd},\;\;\
an_x+(Np-1)bn'_x=an_y+(Np-1)bn'_y=0.
\end{equation}
From the above equation, we have
$\frac{n'_x}{n'_y}=\frac{n_x}{n_y}$ or $\cot \phi'=\cot \phi$ so
that $\phi'=\pi+\phi$. In the other hand, we have the condition
$\beta=0$ for the case that $\rho_1$ is in the subspace spanned by
$J_x$ and $J_y$ (invariant subspace under the rotations about the
$z$ axes), i.e., for the case that we have $n_z=0$. Then,
Eq.(\ref{x3}) implies that $n'_z=\cos \theta'=0$, or
$\theta'=\pi/2$. Therefore, we have
$n'_x=\sin\theta'\cos\phi'=-\cos \phi$ and Eq.(\ref{x3}) leads to
\begin{equation}\label{x4}
p_{_{opt}}=\frac{1}{N}(1+ \frac{a\sin \theta}{b}).
\end{equation}
By substituting $b=\frac{1}{d-1}$, the optimal success probability
is given by
\begin{equation}\label{x5}
p_{opt}=\frac{1}{N}[1+ a(d-1)\sin \theta].
\end{equation}

As an another example, let us consider the equiprobable qubit
states $\rho_{1},\rho_{2},\ldots,\rho_{N}$ with the corresponding
Bloch vectors
\begin{equation}\label{e46}
\textbf{a}_{j}=(a\sin\theta\cos\phi_{j},a\sin\theta\sin\phi_{j},a\cos\theta),\quad
j=1,\ldots,N,
\end{equation}
which share a common latitude of a ball with the radius equal to
$a$. Except for $\theta=\pi/2$, the states are reducible, i.e.,
none of the states are invariant under the rotations about $z$-
axis. By choosing $\phi_1=0$, we have
$$\rho_j=U_j\rho_1U^{-1}_j, \;\ \mbox \;\ U_j=\left(\begin{array}{cc}
                                             e^{-i\phi_j/2} & 0 \\
                                                0 & e^{i\phi_j/2} \\
                                              \end{array}\right)=e^{-i\phi_j\sigma_z/2},\;\;\ j=2,\ldots, N.$$
A it is seen, the operators $U_j$, for $j=1,2,\ldots, N$ generate
the infinite rotation group about $z$-axis, i.e., we choose finite
number of non symmetric states from infinite states- which can be
produce in this way- and discriminate them with minimum error
probability. In the qubit case, $\tau_j$'s are pure and so can be
taken as $\tau_j=1/2(I+{\mathbf{b}}_j.\sigma)$ with
$|{\mathbf{b}}_j|=1$. Similar to the previous example, by
substituting $\mathcal{M}=\alpha I_2+\beta \sigma_z$ and using the
identity (\ref{e100'}), one can obtain the optimal success
probability of discrimination as follows
$$p_{opt}=\frac{1}{N}(1+a\sin \theta),$$
which is special case $d=2 $ of the result (\ref{x5}). This
example has solved in Ref. \cite{hels.maz} via the Helstrom family
of ensemble and using the convex optimization method.\\\\
\textbf{MED between group covariant states}\\ In the special case
in which the states $\rho_i$ are labeled by all of the elements of
a group $G$ (denoted by $\rho_g$) called also group covariant
states or symmetric states, the set of optimal POM $\{\Pi_g, g\in
G\}$ is determined uniquely in terms of $\{\Pi'_g, g\in G\}$. To
see this, assume that the states $\rho_g$ are defined as
\begin{equation}\label{rog}\rho_{g}=U(g)\rho_e U(g)^{-1} ,\;\;\ \tau_{g}=U(g)\tau_e U(g)^{-1},\;\;\ \forall \;\ g\in G
\end{equation}
where, $U$ is an irreducible unitary representation of $G$. These
states are sometimes called group covariant quantum states or
symmetric states.

It should be noticed that, the irreducible representation
associated with the group elements belonging to the center (the
center of a group is defined as a set of elements commuting with
all of the group elements, i.e., $Z=\{g'\in G ; gg'=g'g \;\
\forall g\in G\}$) is multiple of identity matrix, that is we have
$U(g)=e^{i\phi(g)}I$ for all $g\in Z$. Therefore, one can consider
the groups $G$ with trivial center $Z=\{e\}$ or- in the case of
groups with non-trivial center- the quotient group $G/Z$ instead
of $G$ and parameterize the initial states $\rho_i$ and POVM set
$\Pi'_i$, with elements of $G/Z$.

In order to see that, in this special case, the coefficients
$\lambda_g$ in (\ref{p'}) are determined uniquely, we denote
$\sum_{g\in G}\Pi'_g$ by $\Pi$. Then, one can write
$$U(g')\Pi U(g')^{-1}=\sum_{g\in G}U(g')\Pi'(g)U(g')^{-1}=\sum_{g\in G}U(g'g)\Pi'_1U_(g'g)^{-1}=$$$$\sum_{g''\in G}U(g'')\Pi'_1U_(g'')^{-1}=\sum_{g''\in G}\Pi'_{g''}=\Pi, \;\;\ \forall \;\ g'\in G.$$ Due to the ineducability of group representation $U$, the
Schur's lemma implies that $\Pi$ must be multiple of identity
operator, i.e., we have
$$\Pi=\sum_{g\in G}\Pi'_g=\alpha I.$$
Taking the trace of the both sides and using (\ref{p''}), we
obtain
$$|G|Tr \Pi'_e=\alpha d, \;\ \rightarrow \;\ \alpha=|G|.$$
Therefore, we have $\frac{1}{|G|}\sum_{g\in G}\Pi'_g=I$ so that
the set of optimal POVM is given by
$\Pi_g=\lambda_g\Pi'_g=\frac{1}{|G|}\Pi'_g$.\\\\
\textbf{Conclusion}\\
The minimum error discrimination (MED) between quantum mixed
states was studied where, it was shown that the known necessary
and sufficient conditions for MED \cite{Barnett1} are fulfilled by
a Helstrom family of ensembles. Then, a more convenient and
suitable identity as a convex combination of the initial states -
which we try to discriminate between them- and their corresponding
conjugate states was extracted from the MED conditions in order to
obtain optimal set of measurements and corresponding optimal
success probability of discrimination. The introduced identity was
applied for MED between the similarity transformed quantum states.
Finally, as a special case, MED between the group covariant or
symmetric states was considered.


\begin{thebibliography}{99}
\bibitem{Helstrom} C. W. Helstrom, {\it Quantum Detection and Estimation theory}, New York: Academic, (1976).
\bibitem{Holevo} A. Holevo, {\it Probabilistic and Statistical Aspects of Quantum Theory}, Amsterdam: North-Holland, (1982).
\bibitem{Eldar1} Y. C. Eldar, A. Megretski, and G. C. Verghese, IEEE Trans. Inform. Theory 49, 1007
(2003).
\bibitem{Eldar2} Y. C. Eldar, A. Megretski, and G. C. Verghese, IEEE Trans. Inform. Theory 50, 1198
(2004).
\bibitem{Barnett1} S. M. Barnett and S. Croke, J. Phys. A: Math. Theor. 42,
062001, (2009).
\bibitem{Hunter} K. Hunter, in Proceedings of The
Seventh International Conference on Quantum Communication,
Measurement and Computing (QCMC04), Vol. 734 of American Institute
of Physics Conference Series (AIP, 2004), 83–86.
\bibitem{Yuen} Yuen, H., Kennedy, R., and Lax, M., IEEE Trans. Inf. Theory,
IT-21, 125 (1975).
\bibitem{Hwang} W. Y. Hwang and J. Bae, J. Math. Phys. 51, 022202, (2010).
\bibitem{Assalini} A. Assalini, G. Cariolaro, and G. Pierobonar, Phys. Rev. A 81, 012315, (2010).
\bibitem{Barnett2} S. M. Barnett, Phys. Rev. A 64, 030303(R),
(2001).
\bibitem{Chou} C. L. Chou and L. Y. Hsu, Phys. Rev. A 68, 042305,
(2003).
\bibitem{Barnett3} S. M. Barnett and S. Croke, Adv. Opt. Photon. 1, 238, (2009).
\bibitem{Kimura} G. Kimura, T. Miyadera, and H. Imai, Phys. Rev. A 79, 062306, (2009).
\bibitem{hels.maz} M.A. Jafarizadeh, Y. Mazhari, M. Aali, Quantum Inf
Process, DOI 10.1007/s11128-010-0185-y, (2010).
\bibitem{joshi} A. W. Joshi, {\it Elements of group theory for
physicists}, New Age International (P) Limited, Publishers,
(1997).
\end{thebibliography}
\end{document}